\documentclass{article}

\usepackage{epsfig}
\usepackage{amssymb,latexsym}

\newcommand{\beq}{\begin{equation}}
\newcommand{\eeq}{\end{equation}}
\newcommand{\bqa}{\begin{eqnarray}}
\newcommand{\eqa}{\end{eqnarray}}
\newcommand{\fr}{\frac}

\begin{document}
\title{\bf \Large Strength and genericity of singularities in Tolman-Bondi-de Sitter collapse}
\author{{\large S\' ergio M. C. V. Gon\c calves\footnote{E-mail: sergiog@its.caltech.edu, Tel: +1-626-395-8753, Fax: +1-626-796-5675}} \\
{\em \large Theoretical Astrophysics, California Institute of Technology} \\
{\em \large Pasadena, California 91125, U.S.A.}}
\date{\small \today}

\maketitle
\begin{abstract}
We study the curvature strength and visibility of the central singularity arising in Tolman-Bondi-de Sitter collapse. We find that the singularity is visible and Tipler strong along an infinite number of timelike geodesics, independently of the initial data, and thus stable against perturbations of the latter. \\
\\
PACS: 04.20.Dw \hspace{0.5cm} KEYWORDS: black hole, cosmic censorship, singularity.
\end{abstract}
\newpage
\section{Introduction}

It has recently been shown by Gon\c calves \cite{goncalves00a} and, independently, by Deshingkar et al. \cite{deshingkaretal00}, that a central curvature singularity develops in asymptotically de Sitter spherical dust collapse. When curvature growth is examined along radial null directions, the singularity is found to be visible, provided the initial central density distribution, $\rho_{\rm c}(r)$, obeys certain differentiability criteria \cite{goncalves00a}. Let $n\in{\mathbb N}^{+}$ be such that $(\partial^{n}\rho_{c}(r)/\partial r^{n})_{r=0}\equiv\rho_{n}$ is the first non-vanishing derivative of the central energy density distribution. The singularity is visible for $n\leq3$. For the cases $n\leq2$, a single radial null geodesic is emitted from the singularity; for $n=3$, if, in addition, $\rho_{3}$ obeys certain constraints, an entire family of outgoing radial null geodesics has its past endpoint at the singularity. Regarding curvature strength, the singularity is Tipler strong \cite{tipler77} for $n=3$, and may or may not be Tipler strong for $n\leq2$. This behavior is analogous to the one found in asymptotically flat spherical dust collapse, whose analysis along radial null directions shows the central singularity to be visible for $n\leq3$ \cite{joshi&dwivedi93,dwivedi&joshi97,jhingan&joshi97}, and gravitationally strong for $n=3$ (weak for $n\leq2$) \cite{jhingan&joshi97}.

All the previous results were derived based on an analysis of curvature growth along radial {\em null} geodesics. However, such analyses, based solely on radial null geodesics, are too restrictive and have given rise to the misleading notion that the physical nature of the singularity in asymptotically flat dust collapse is critically determined by the initial data. A notable exception is the work of Deshingkar, Joshi and Dwivedi \cite{deshingkar&joshi&dwivedi99}; as their analysis shows, when one considers {\em timelike} radial geodesics, the singularity is found to be locally naked and Tipler strong for an infinite number of non-spacelike radial geodesics, irrespective of the initial data. A recent detailed analysis of non-spacelike geodesics in (asymptotically flat) dust collapse by Deshingkar and Joshi \cite{deshingkar&joshi00} showed that a family of radial null and timelike outgoing geodesics has its past endpoint at the singularity, thereby confirming the results of \cite{deshingkar&joshi&dwivedi99}.

In this paper, we perform an analysis of visibility and curvature growth along radial timelike directions for the case of Tolman-Bondi-de Sitter  (TBdS)  collapse. We also find that the central curvature singularity is at least locally naked, and Tipler strong, independently of the initial data, which implies stability against perturbations of the latter. Such a behavior is not unexpected, since the structure of the singularity in TBdS collapse is quite similar to that of asymptotically flat dust collapse \cite{goncalves00a,deshingkaretal00,deshingkar&joshi00,joshi&dwivedi99,barveetal99}. This result adds further evidence to the robustness of dust collapse against the introduction of a cosmological constant.

Geometrized units, in which $G=c=1$, are used throughout.

\section{Central naked singularity in Tolman-Bondi-de Sitter collapse}

The Tolman-Bondi-de Sitter family of solutions is given by a spherically symmetric metric, written here in comoving coordinates $\{t,r,\theta,\phi\}$:
\beq
ds^{2}=-dt^{2}+\fr{R'^{2}}{1-k}dr^{2}+R^{2}d\Omega^{2}, \label{tbm}
\eeq
where $'\equiv\partial_{r}$, together with the stress-energy tensor,
\beq
T_{ab}=\rho(t,r)\delta_{a}^{t}\delta_{b}^{t}-\fr{\Lambda}{8\pi}g_{ab},
\eeq
where $\rho(t,r)$ is the energy density distribution, and $\Lambda>0$ the cosmological constant. In spherical symmetry there are no gravitational degrees of freedom, and the metric is thus uniquely determined from the initial data, $\rho(0,r)$ and $\dot{R}(0,r)$, where $\dot{}\equiv\partial_{t}$ and $R(t,r)$ is a solution of 
\beq
\dot{R}^{2}=\fr{2m}{R}-k+\fr{\Lambda}{3}R^{2}, \label{tb1}
\eeq
with initial data given by
\beq
m'(r)=4\pi R^{2}R'\rho(t,r), \label{mtb}
\eeq
where $m(r)$ is the Misner-Sharp mass ~\cite{misner&sharp64} (which agrees with Hawking's quasi-local mass ~\cite{hawking} and ADM mass, in the appropriate limits) and is uniquely determined from the initial density profile, $\rho(0,r)$.

The case $k=0$ (corresponding to gravitationally unbound matter configurations) is qualitatively equivalent to those with $k\neq0$, and it allows for an analytical solution of Eq. (\ref{tb1}) in closed form:
\bqa
R(t,r)&=&\left(\fr{6m}{\Lambda}\right)^{\fr{1}{3}}\sinh^{2/3}T(t,r), \label{rad} \\
T(t,r)&\equiv&\fr{\sqrt{3\Lambda}}{2}\left[t_{\rm c}(r)-t\right], \label{time}
\eqa
where $t_{\rm c}(r)$ is the proper time for complete collapse of a shell with initial area radius $R(0,r)$, which is fixed by Eq. (\ref{rad}) at $t=0$:
\beq
t_{\rm c}(r)=\fr{2}{\sqrt{3\Lambda}}\sinh^{-1}\left(\sqrt{\fr{\Lambda r^{3}}{6m}}\right),
\eeq
where the scaling $R(0,r)=r$ was adopted. 

The relevant derivatives of the area radius are
\bqa
R'(t,r)&=&R\left(\fr{m'}{3m}+\sqrt{\fr{\Lambda}{3}}t'_{\rm c}\coth T\right), \label{Rpr} \\
\dot{R}(t,r)&=&-\sqrt{\fr{\Lambda}{3}}R\coth T. \label{Rdot}
\eqa

At $t=t_{\rm c}$, $R(t_{\rm c},0)=0$, and the Kretschmann scalar 
\beq
{\mathcal K}\equiv R_{abcd}R^{abcd}=3\fr{m'}{R^{4}R'^{2}}-8\fr{mm'}{R^{5}R'}+12\fr{m'^{2}}{R^{6}}+\fr{8}{3}\Lambda(\Lambda+4\pi\rho),
\eeq
diverges, thereby signaling the existence of a curvature singularity at $r=0$. This singularity will be at least locally naked if the outgoing radial null geodesics equation,
\beq
\fr{dt}{dr}=R', \label{ng}
\eeq
admits a regular solution, for given initial data, at $t=t_{\rm c}(0)=t_{0}$. Expanding $\rho(0,r)\equiv \rho_{\rm c}(r)$  near $r=0$,
\beq
\rho_{\rm c}(r)=\sum_{i=0}^{+\infty} \rho_{i}r^{i},
\eeq
near the singularity (where $r,T\rightarrow0^{+}$) we have, to leading order,
\bqa
m(r)&=&m_{0}r^{3}+m_{n}r^{n+3}+{\mathcal O}(r^{n+4}), \label{mass} \\
t_{\rm c}(r)&=&t_{0}+t_{n}r^{n}+{\mathcal O}(r^{n+1}), \label{tcoll} \\
R(t,r)&=&\left(\fr{9}{2}\right)^{\fr{1}{3}}\left(m_{0}^{\fr{1}{3}}r+M_{n}r^{n+1}\right) \left(t_{0}+t_{n}r^{n}-t\right)^{\fr{2}{3}}+\,{\mathcal O}(r^{n+2})\times{\mathcal O}(T^{\fr{8}{3}}), \label{bigr}
\eqa
where $t_{n}$ and $M_{n}$ are real coefficients linear in $m_{n}=(4\pi/n)\rho_{n}$, with $n>0$; $\rho_{n}\equiv(\partial^{n}\rho_{c}/\partial r^{n})_{r=0}$ is the first non-vanishing derivative of the central energy density distribution, and
\beq
t_{0}=\sinh^{-1}\left(\sqrt{\fr{\Lambda}{6m_{0}}}\right). \label{tzero}
\eeq

Assuming that there is a regular solution to the outgoing null geodesics equation near $r=0$, which we write, to leading order, as
\beq
t=t_{0}+ar^{\sigma}, \label{null}
\eeq
where $a,\sigma\in\mathbb{R}^{+}$, the singularity will be at least locally naked if Eq. (\ref{ng}) admits a self-consistent solution, and covered otherwise. Since the geodesic must lie on the spacetime, from Eqs. (\ref{tcoll}) and (\ref{null}) it follows that $\sigma\geq n$. From Eqs. (\ref{bigr}) and (\ref{null}), we obtain
\beq
R(t,r)=\left(\fr{9m_{0}}{2}\right)^{\fr{1}{3}}t_{n}^{\fr{2}{3}}r^{\fr{2n}{3}+1}+{\mathcal O}(r^{\sigma+2-\fr{n}{3}}). \label{newr}
\eeq
Let us first consider the $\sigma>n$ case. Differentiating Eqs. (\ref{null}) and (\ref{newr}) w.r.t. $r$, and requiring a self-consistent solution of Eq. (\ref{ng}) yields 
\bqa
\sigma&=&1+\fr{2n}{3}, \label{slope} \\
a&=&\left(\fr{9m_{0}}{2}\right)^{\fr{1}{3}}t_{n}^{\fr{2}{3}}. 
\eqa
The condition $\sigma>n$ now reads $n<3$.  For $n=1,2$ (i.e., for $\rho_{1}\neq0$, or $\rho_{1}=0$ and $\rho_{2}\neq0$) there is a self-consistent solution to the outgoing radial null geodesics equation in the limit $t\rightarrow t_{0}$, $r\rightarrow0$, and thus there is at least one outgoing radial null geodesic starting from the singularity.

For $n=\sigma=3$, an outgoing radial null geodesic exists, provided \cite{goncalves00a}
\beq
\fr{M^{2}}{2}-4+\sqrt{M^{4}+16-4M^{2}}<t_{3}<\fr{1}{8}M^{4} \label{t3c}
\eeq
where $M\equiv(9m_{0}/2)^{\fr{1}{3}}$. For the special case (with $n=3$) when $M>8.799799016$ and $\rho_{3}$ is uniquely determined from $\rho_{0}$ [cf. Eq. (45) in \cite{goncalves00a}], an entire one-parameter family of outgoing radial null geodesics departs from the singularity. Clearly, this is a set of measure zero in the initial data.

\section{Visibility}

Let us consider radial timelike geodesics (RTGs), with tangent vector $K^{a}=\fr{dx^{a}}{d\tau}$, where $\tau$ is an affine parameter along the geodesic, and
\bqa
&&K^{t}=\pm\sqrt{1+R'^{2}(K^{r})^{2}}, \label{ge1} \\
&&\dot{K^{r}}R'+2K^{r}\dot{R}'+\fr{K^{r}}{K^{t}} (K^{r})'R'+\fr{(K^{r})^{2}}{K^{t}}R''=0, \label{ge}
\eqa
where the first equation is simply $K^{a}K_{a}=-1$, and the second follows from the geodesic equation. By inspection, one sees that Eq. (\ref{ge}) admits the trivial solution
\bqa
K^{t}&=&\pm1, \label{kt} \\
K^{r}&=&0, \label{kr}
\eqa
which leads to
\bqa
t=t_{0}\pm(\tau-\tau_{0}), \label{timet} \\
r=r_{0}=\mbox{const.},
\eqa
where the plus or minus sign refers to outgoing or ingoing RTGs, respectively. The outgoing RTG departing from the singularity is given
by $r=0$, and $t=t_{0}+\tau-\tau_{0}$, and thus does not
belong in the spacetime. The ingoing RTG is given by $r=0$,
$t=t_{0}-\tau+\tau_{0}$, where $t_{0}=t_{\rm c}(0)=0$ is the time at which the RTG arrives at the singularity.

\subsection{Genericity}

Since Eq. (\ref{ge}) is a mixed first-order linear PDE for $K^{r}(t,r)$, its solution need not be unique \cite{kevorkian99}. Indeed, one can explicitly show that there are infinitely many other solutions, as follows. Near the singularity, we can write, to leading order
\bqa
t_{\rm RTG}(r)&=&t_{0}+br^{p}, \label{rtg} \\
R'&=&a_{1}r^{q}, \\
\dot{R}'&=&a_{2}r^{q-p}, \\
R''&=&a_{3}r^{q-1},
\eqa
where $b,p,q\in{\mathbb R}^{+}$, and $a_{i}\in{\mathbb R}\;(i=1,2,3)$ are constants uniquely determined from the initial data, and the $t$ dependence of $R$ has been absorbed in the exponent $q$ [which can be done without loss of generality, since $R$ is evaluated {\em along} the geodesic defined by Eq. (\ref{rtg})].

Let us now assume that $K^{r}(t,r)\propto (t-t_{0})^{\alpha}r^{\beta}$, where $\alpha,\beta\in {\mathbb R}$. From Eq. (\ref{rtg}) we have then
\bqa
K^{r}&=&c_{1}r^{\alpha p+\beta}, \\
\dot{K}^{r}&=&c_{2}r^{(\alpha-1)p+\beta}, \\
(K^{r})'&=&c_{3}r^{\alpha p+\beta-1},
\eqa
where $c_{i}\in{\mathbb R}\;(i=1,2,3)$, are free constants, yet to be determined. With these ansatze, Eqs. (\ref{ge1})-(\ref{ge}) read:
\bqa
&&K^{t}=\pm\sqrt{1+a_{1}^{2}c_{1}^{2}r^{2(q+\alpha p+\beta)}}, \\
&&a_{1}c_{2}r^{(\alpha-1)p+\beta-q}+2a_{2}c_{1}r^{(\alpha-1)p+\beta-q}+\fr{a_{1}c_{1}c_{3}}{K^{t}}r^{2(\alpha p+\beta)+q-1}+\fr{a_{3}c_{1}^{2}}{K^{t}}r^{2(\alpha p+\beta)+q-1}=0. \label{ge2}
\eqa
If the free constants $c_{i}$ and $p,q,\alpha,\beta$ are such that the system (\ref{ge1})-(\ref{ge}) is self-consistent, then there are outgoing radial timelike geodesics emanating from the singularity---each one of them uniquely determined by a particular set $\{p,q,\alpha,\beta, c_{i}, i=1,2,3\}$. We now look for solutions of the form $K^{t}=\mbox{const.}$, and in which the left-hand-side of Eq. (\ref{ge2}) is homogeneous in $r$. This implies
\bqa
&&\alpha p+q+\beta=0, \\
&&\alpha p+q+\beta-p=2(\alpha p+\beta)-1+q,
\eqa
which is solved by
\beq
p=1+q=\fr{1-\beta}{1+\alpha}.
\eeq
Equation (\ref{ge}) becomes then
\beq
C(c_{i})r^{\fr{\beta-1}{1+\alpha}}=0,
\eeq
which has an {\em infinite} number of solutions, given by the algebraic constraint
\beq
C(c_{i})=a_{1}c_{2}+2a_{2}c_{1}\pm\fr{a_{1}c_{1}c_{3}+a_{3}c_{1}^{2}}{\sqrt{1+a_{1}^{2}c_{1}^{2}}}=0.
\eeq
Hence, there are infinitely many radial outgoing (ingoing) timelike geodesics, with past (future) endpoint at the singularity. The existence of these families of RTGs is independent of the initial data---which enters the algebraic constraint via the $a_{i}$ constant coefficients---since the algebraic constraint $C(c_{i})=0$ always has a solution, due to the three degrees of freedom for the $c_{i}$ coefficients.

One can also prove the existence of an infinite number of timelike geodesics with endpoint at the singularity, by resorting to simple causal structure considerations, as follows.

The existence of a future-directed timelike geodesic
$\sigma(\lambda)$ with future endpoint at the singularity, implies that its chronological past $I^{-}[\sigma]$ is a {\em
timelike indecomposable past set} (TIP) \cite{gerochetal72}. In other
words, the TIP is generated by $\sigma$, which is timelike and
future-inextendible. The locally naked singularity itself---which is
the future endpoint of $\sigma$, $\sigma(\lambda_{0})$---constitutes a
singular TIP which contains the past $I^{-}(p)$ of any point
$p\in\sigma\backslash\{\sigma(\tau_{0})\}$ \cite{penrose78}. Consider
now another point $q\not\in\sigma$ in the past of the singularity and
in the chronological past of $p$. Since $I^{-}(q)\subset I^{-}(p)$, it
follows that there exists a timelike curve from $p$ to $q$, say
$\zeta$, satisfying $I^{-}[\zeta]\subset I^{-}[\sigma]$. Consider then
a small compact ball ${\mathcal B}$---with a suitably defined
``radius'', e.g., proper geodesic distance along $\sigma$,
$d_{\sigma}(\lambda,\lambda_{0})$---in the neighborhood of
$\sigma(\tau_{0})$, partially contained in TIP $I^{-}[\sigma]$ and in
the chronological future of $p$, i.e., ${\mathcal B}\cap
I^{-}[\sigma]={\mathcal C}\nsubseteq\emptyset$ and $I^{+}[{\mathcal
C}]\subset I^{+}(p)$. For each point $x\in{\mathcal C}$, there is a
timelike geodesic from $p$ to $x$. Since ${\mathcal C}$ is compact,
there is an infinite number of such points that can be joined by
timelike geodesics from $p\in\sigma$. Now, let $p$ be an arbitrary
point on $\sigma$ and take the limit of approach to the singularity,
$d_{\sigma}(\lambda,\lambda_{0})\rightarrow0^{+}$; it then follows
that there is an infinite number of future-inextendible timelike
geodesics with future endpoint at the singularity.

\subsection{Apparent horizon}

The existence of outgoing timelike geodesics with past endpoint at the singularity renders the latter visible if and only if the geodesics are emitted before or at the time of formation of the apparent horizon (AH). The AH is a spacelike 2-surface defined by the outer boundary of a trapped region, and in the adopted coordinates it is given by $R_{,a}R_{,b}g^{ab}=0$. With the metric (\ref{tbm}) and Eq. (\ref{tb1}) we have then
\beq
\fr{\Lambda}{3}R^{3}-R+2m=0.
\eeq
This equation has three distinct real roots if $3m\sqrt{\Lambda}<1$, two of which are positive and given by
\beq
R_{\pm}=\fr{2}{\sqrt{\Lambda}}\sin\left[\fr{1}{3}\sin^{-1}(3m\sqrt{\Lambda})+\fr{2\pi}{3}\delta(1\pm1)\right], \label{rbh}
\eeq
with $R_{-}>R_{+}>0$, corresponding to the choice $0\leq\sin^{-1}\omega\leq\pi/2$, $0\leq\omega\leq1$. The third root, $R_{3}=-R_{-}-R_{+}$, is negative and hence unphysical. $R_{-}$ is a generalized cosmological horizon ($R_{-}=\sqrt{3/\Lambda}$, when $m=0$) and $R_{+}$ the black hole apparent horizon ($R_{+}=2m$ when $\Lambda=0$; the apparent and event horizons coincide in the static case). For $3m\sqrt{\Lambda}=1$, the two horizons coincide. If $3m\sqrt{\Lambda}>1$, there is one negative real root and two complex (conjugate) roots, all of which are unphysical: the spacetime does not admit any horizons in this case. From the ``+'' solution in Eq. (\ref{rbh}), together with Eqs. (\ref{rad})-(\ref{time}), we obtain
\beq
t_{\rm AH}(r)=t_{\rm c}(r)-\fr{2}{\sqrt{\Lambda}}\sinh^{-1}\left\{\sqrt{\fr{8}{3m\Lambda}}\left[ \sin\left(\fr{1}{3}\sin^{-1}(3m\sqrt{\Lambda})\right)\right]^{3/2}\right\}. \label{tah}
\eeq
At the origin we have $t_{\rm AH}(0)=t_{\rm c}(0)-\xi$, where $0\leq \xi<(2/\sqrt{3\Lambda})\sinh^{-1}\sqrt{\sqrt{3}/(m\Lambda)}$. Hence, $t_{\rm ORTG}(0)\geq t_{\rm AH}(0)$. If $t_{\rm ORTG}(0)>t_{\rm AH}(0)$, any radial geodesic emitted from the singularity at $r=0$ will do so {\em after} the AH forms, and is therefore unavoidably trapped---the singularity is covered. If $t_{\rm ORTG}(0)=t_{\rm AH}(0)$, the singularity is at least locally naked, provided $\left(\fr{dt_{\rm AH}}{dr}\right)_{r=0}>\left(\fr{dt_{\rm ORTG}}{dr}\right)_{r=0}$. For $\xi=0$, $\fr{dt_{\rm AH}}{dr}=\fr{dt_{\rm c}}{dr}>0$ (i.e., there are no shell-crossing singularities); if $p\neq1$ in Eq. (\ref{rtg}), then $\left(\fr{dt_{\rm ORTG}}{dr}\right)_{r=0}=0$, and the singularity is thus always visible. For the special case $p=1$, appropriate choice for $b$ in Eq. (\ref{rtg}) renders the singularity visible.

\section{Curvature strength}

Let $\gamma(\tau)$ be a RTG, with tangent vector $K^{a}=\fr{dx^{a}}{d\tau}$, that terminates at the singularity at $\tau=\tau_{0}$. Such a singularity is said to be Tipler strong along $\gamma(\tau)$ if the volume three-form $V(\tau)$ vanishes in the limit $\tau\rightarrow\tau_{0}$ \cite{tipler77}. If the scalar $\Psi\equiv R_{ab}K^{a}K^{b}$ obeys the {\em strong limiting focusing condition},
\beq
\lim_{\tau\rightarrow\tau_{0}} (\tau-\tau_{0})^{2}\Psi>0,
\eeq
then the singularity is gravitationally strong in the sense of Tipler \cite{clarke&krolak85}. This sufficient condition guarantees that any three-form defined along $\gamma(\tau)$ vanishes in the limit $\tau\rightarrow\tau_{0}$, due to unbounded curvature growth.

From Eqs. (\ref{tbm})-(\ref{tb1}) and (\ref{kt})-(\ref{kr}), we have,
\beq
\Psi=\fr{m'}{R^{2}R'}-\Lambda.
\eeq
Using Eqs. (\ref{mass})-(\ref{bigr}) and (\ref{timet}), we obtain, near $r=0$,
\bqa
\Psi&=&\fr{2}{3}\eta^{-\fr{4}{3}}\left(\eta^{\fr{2}{3}}+\fr{2}{3}nt_{n}r^{n}\eta^{-\fr{1}{3}}\right)^{-1}, \\
\eta&\equiv&t_{n}r^{n}\mp(\tau-\tau_{0}).
\eqa
Hence,
\beq
\lim_{\tau\rightarrow\tau_{0}} (\tau-\tau_{0})^{2}\Psi=\fr{2}{3},
\eeq
and the singularity is therefore Tipler strong. This is a rather robust result, in that it is independent of the initial data and it holds true along an infinite number of timelike geodesics. We note that, since all the sufficient criteria for Tipler strong singularities are sufficient for  {\em deformationally strong} ones \cite{nolan99,ori00}, the central curvature singularity in TBdS collapse is also deformationally strong.

\section{Concluding remarks}

We have shown that the central singularity arising in spherical dust collapse with a positive cosmological constant is Tipler strong and at least locally naked, along an infinite number of timelike geodesics, independently of the initial data. Such a singularity could possibly constitute a counter-example to the strong cosmic censorship conjecture \cite{penrose78,penrose79}, if it were to remain stable against perturbations of matter and geometry. The work of Harada et al. \cite{haradaetal98} and Iguchi et al. \cite{iguchietal99,iguchietal00}, showed that the central curvature singularity in asymptotically flat dust collapse is marginally stable against non-spherical linear perturbations. These authors examined the dynamical evolution of non-spherical metric, matter, and matter coupled to matter perturbations in Lema\^{\i}tre-Tolman-Bondi collapse. They found that whilst the singularity itself is not a strong source of gravitational radiation, the non-spherical perturbations remain well-behaved and finite in the vicinity of the singularity and in the limit of approach to the Cauchy horizon. Given the similar structure of the central singularity in Lema\^{\i}tre-Tolman-Bondi collapse and the one discussed in this paper, one may speculate that similar stability properties also hold when $\Lambda>0$. This issue is currently being investigated.

\section*{Acknowledgments}
I thank Sanjay Jhingan and Kip Thorne for useful discussions. This work was supported by FCT (Portugal) Grant PRAXIS XXI-BPD-16301-98, and by NSF Grant AST-9731698.


\begin{thebibliography}{99}

\bibitem{goncalves00a}
S. M. C. V. Gon\c calves, Phys. Rev. D {\bf 63}, 064017 (2001).

\bibitem{deshingkaretal00}
S. S. Deshingkar, A. Chamorro, S. Jhingan, and P. S. Joshi, {\em ``Gravitational collapse and cosmological constant''}, gr-qc/0010027.

\bibitem{tipler77}
F. J. Tipler, Phys. Lett. {\bf 64A}, 8 (1977).

\bibitem{joshi&dwivedi93}
P. S. Joshi and I. H. Dwivedi, Phys. Rev. D {\bf 47}, 5357 (1993).

\bibitem{dwivedi&joshi97}
I. H. Dwivedi and P. S. Joshi, Class. Quantum Grav. {\bf 14}, 1223 (1997).

\bibitem{jhingan&joshi97}
S. Jhingan and P. S. Joshi, Ann. Isr. Phys. Soc. {\bf 13}, 357 (1997).

\bibitem{deshingkar&joshi&dwivedi99}
S. S. Deshingkar, P. S. Joshi, and I. H. Dwivedi, Phys. Rev. D {\bf59}, 044018 (1999).

\bibitem{deshingkar&joshi00}
S. S. Deshingkar and P. S. Joshi, Phys. Rev. D {\bf 63}, 024007 (2000).

\bibitem{joshi&dwivedi99}
P. S. Joshi and I. H. Dwivedi, Class. Quantum Grav. {\bf 16}, 41 (1999).

\bibitem{barveetal99}
S. Barve, T. P. Singh, C. Vaz, and L. Witten, Class. Quantum Grav. {\bf 16}, 1727 (1999).

\bibitem{misner&sharp64}
C. W. Misner and D. H. Sharp, Phys. Rev. B {\bf 136}, 571  (1964).

\bibitem{hawking}
S. W. Hawking, J. Math. Phys. {\bf 9}, 598 (1968).

\bibitem{kevorkian99}
J. Kevorkian, {\em Partial Differential Equations: Analytical Solution Techniques} (Springer-Verlag, New York, 1999).

\bibitem{gerochetal72}
R. Geroch, E. H. Kronheimer, and R. Penrose, Proc. R. Soc. Lond. {\bf A327}, 545 (1972).

\bibitem{penrose78}
R. Penrose, in {\em Theoretical Principles in Astrophysics and Relativity}, Eds. N. R. Liebowitz, W. H. Reid, and P. O. Vandervoort (Chicago University Press, Chicago, 1978).

\bibitem{penrose79}
R. Penrose, in {\em General Relativity, An Einstein Centenary Survey}, Eds. S. W. Hawking and W. Israel (Cambridge University Press, Cambridge, England, 1979).

\bibitem{clarke&krolak85}
C. J. S. Clarke, and A. Kr\' olak, J. Geom. Phys. {\bf 2}, 127 (1985).

\bibitem{nolan99}
B. C. Nolan, Phys. Rev. D {\bf 60}, 0124014 (1999).

\bibitem{ori00}
A. Ori, Phys. Rev. D {\bf 61}, 064016 (2000).

\bibitem{haradaetal98}
T. Harada, H. Iguchi, and K. Nakao, Phys. Rev. D {\bf 58}, 041502 (1998).

\bibitem{iguchietal99}
H. Iguchi, T. Harada, and K. Nakao, Prog. Theor. Phys. {\bf 101}, 1235 (1999).

\bibitem{iguchietal00}
H. Iguchi, T. Harada, and K. Nakao, Prog. Theor. Phys. {\bf 103}, 53 (2000).

\end{thebibliography}
\end{document}